\documentclass[reprint,noshowpacs,noshowkeys,prd,balancelastpage,nofootinbib]{revtex4-1}
\usepackage[utf8]{inputenc}
\usepackage{color}
\usepackage{amsmath}
\usepackage{amssymb}
\usepackage{graphicx}
\usepackage[unicode=true,
 bookmarks=false,
 breaklinks=false,pdfborder={0 0 1},backref=section,colorlinks=true]
 {hyperref}
\hypersetup{
 linkcolor=blue,urlcolor=blue,citecolor=blue}

\makeatletter

\usepackage{amsfonts}\usepackage{latexsym}
\usepackage{dcolumn}
\usepackage{epsfig}\usepackage{amsthm}
\makeatother

\begin{document}
\title{Behavior of a Free Quantum Particle in the Poincaré Upper Half-Plane
Geometry}
\author{Parham Dehghani}
\email{parham.dehghani@emu.edu.tr}

\affiliation{Department of Physics, Faculty of Arts and Sciences, Eastern Mediterranean
University, Famagusta, North Cyprus via Mersin 10, Turkey~\\
 }
\author{S. Danial Forghani}
\email{danial.forghani@final.edu.tr}

\affiliation{Faculty of Engineering, Final International University, Kyrenia, North
Cyprus via Mersin 10, Turkey~\\
 }
\author{S. Habib Mazharimousavi}
\email{habib.mazhari@emu.edu.tr}

\affiliation{Department of Physics, Faculty of Arts and Sciences, Eastern Mediterranean
University, Famagusta, North Cyprus via Mersin 10, Turkey~\\
 }
\begin{abstract}
Inspired by the recent work of Filho \textit{et al.} \cite{Filho},
a Hermitian momentum operator is introduced in a general curved space
with diagonal metric. The modified Hamiltonian associated with this
new momentum is calculated and discussed. Furthermore, granting the
validity of the Heisenberg equation in a curved space, the Ehrenfest
theorem is generalized and interpreted with the new position-dependent
differential operator in a curved space. The modified Hamiltonian
leads to a modified time-independent Schrodinger equation, which is
solved explicitly for a free particle in the Poincare\ upper half-plane
geometry. It is shown that a ``free particle'' does not behave as
it is totally free due to curved background geometry.
\end{abstract}
\date{\today }
\maketitle

\section{Introduction}

Quantum mechanics has been expanded and theorized in the Euclidean
space. It is then an important pursuit to expand the algebraic language
of quantum mechanics within the framework of a curved geometry, say
the geometry of the general theory of relativity. For a long time,
the combination of quantum field theory and general relativity has
been a holy grail that no one has ever managed to reach. Then, any
endeavor to change the existing pillars of quantum world such that
its modified picture lives up to a unification with general relativity
is appreciated. There are thus phenomenological research works addressing
the extension of quantum mechanics on a curved geometry \citep{CCB,chung,park,perivo,dehghani,dehghani1,chung1,fityo,gnatenko}.
Dirac and Klein-Gordon oscillators are examined on Anti-de Sitter
space which include deformed Heisenberg uncertainty principle and
non-commutative momentum space \citep{merad1}. In Ref. \citep{CCB},
quantum mechanics in curved spacetime has been studied. To do so,
the author employed a spherically symmetric static spacetime with
the line element 
\begin{equation}
ds^{2}=-f\left(r\right)dt^{2}+\frac{1}{f\left(r\right)}dr^{2}+r^{2}\left(d\theta^{2}+\sin^{2}\theta\,d\phi^{2}\right)\,,\label{1}
\end{equation}
and the momentum operator is defined in the form 
\begin{equation}
p_{j}=\frac{-i\hbar}{\sqrt{g}}\,\partial_{j}\sqrt{g}\,,\label{2}
\end{equation}
where $j=r,\theta,\phi$ and $g$ is the determinant of the space
part of the metric tensor i.e. $g=det\left(g_{ij}\right)=\frac{1}{f\left(r\right)}r^{4}\sin^{2}\theta$.
Upon this definition, the generalized Laplace operator has been extracted
as 
\begin{equation}
\nabla^{2}=\frac{1}{\sqrt{g}}\,\partial_{j}\left(\sqrt{g}g^{jj}\,\partial_{j}\right)\,.\label{3}
\end{equation}
On the other hand, the energy operator corresponds to the time component
of the metric tensor such that 
\begin{equation}
E=\frac{i\hbar}{\sqrt{f}}\,\partial_{0}\,.\label{4}
\end{equation}
The author used the space-Laplace operator (\ref{3}) and energy operator
(\ref{4}) to construct the Klein-Gordon and Dirac equation for spin
$0$ and $\frac{1}{2}$ particles, respectively. Beside the applications
of these operators mentioned in Ref. \citep{CCB}, one may assume
the spacetime metric (\ref{1}) to be of the form 
\begin{equation}
ds^{2}=-dt^{2}+\frac{1}{f\left(r\right)}dr^{2}+r^{2}\left(d\theta^{2}+\sin^{2}\theta\,d\phi^{2}\right)\,,\label{5}
\end{equation}
which consequently implies (\ref{3}) for the space-Laplace, but the
energy operator (\ref{4}) modifies as 
\begin{equation}
E=i\hbar\,\partial_{0}\,.\label{6}
\end{equation}
Upon considering the time $t$ to be the proper time, one may construct
the curved Schrodinger equation as of 
\begin{equation}
\left(-\frac{\hbar^{2}}{2m}\nabla^{2}+V\left(\overrightarrow{r}\right)\right)\Psi\left(r,t\right)=i\hbar\,\partial_{0}\Psi\left(r,t\right)\,.\label{7}
\end{equation}
This is what basically has been considered in a recent paper \citep{Filho}
by Filho \textit{et al.,} where they have introduced the momentum
operator of a quantum particle moving in a curved space with a diagonal
metric. Such a momentum depends on the structure of the curved space,
and consequently the Heisenberg uncertainty relation gets modified.
Moreover, in Ref. \citep{Filho} the authors have explicitly shown
that the Ehrenfest theorem changes due to the curvature of the space
where the particle's motion is under consideration. They have introduced
a non-additive translation operator in a curved space with a diagonal
metric tensor. In accordance with their approach, in a one-dimensional
curved metric with line element 
\begin{equation}
ds^{2}=g_{xx}\,dx^{2}\,,\label{8}
\end{equation}
an infinitesimal translation from an initial point $x$ to a final
point $x+g_{xx}^{^{-\frac{1}{2}}}dx$, one introduces the translation
operator to be 
\begin{equation}
T\left(dx\right)\left|x\right\rangle =\left|x+g_{xx}^{^{-\frac{1}{2}}}dx\right\rangle \,.\label{9}
\end{equation}
This, however, implies that 
\begin{equation}
T\left(dx\right)=1-i\frac{P_{x}}{\hbar}dx\,,\label{10}
\end{equation}
in which $P_{x}$ is the momentum operator in $x$ direction. In three-dimensional
curved space with a diagonal metric given by 
\begin{equation}
ds^{2}=g_{xx}\,dx^{2}+g_{yy}\,dy^{2}+g_{zz}\,dz^{2}\,,\label{11}
\end{equation}
one writes 
\begin{multline}
T\left(d\overrightarrow{r}\right)\left|x,y,z\right\rangle =\\
\Big|x+g_{xx}^{^{-\frac{1}{2}}}dx,y+g_{yy}^{^{-\frac{1}{2}}}dy,\,z+g_{zz}^{^{-\frac{1}{2}}}dz\Big\rangle\,,\label{12}
\end{multline}
with 
\begin{equation}
T\left(d\overrightarrow{r}\right)=1-i\frac{\overrightarrow{P}}{\hbar}\cdot d\overrightarrow{r}\,.\label{13}
\end{equation}
Hence, Eq. (\ref{10}) or (\ref{13}) admits 
\begin{equation}
\left[x,P_{x}\right]=i\hbar\,g_{xx}^{^{-\frac{1}{2}}}\,,\label{14}
\end{equation}
with similar relations for $y$ and $z$. Therefore, one has to introduce
$P_{j}$ such that the corresponding commutation relation (\ref{14})
holds. The first choice is to set 
\begin{equation}
P_{j}=-i\hbar\,g_{jj}^{^{-\frac{1}{2}}}\,\partial_{j}\,,\label{15}
\end{equation}
where $j$ stands for $x,y$ or $z.$ Here, we would like to add that
$P_{j}$, as given in (\ref{15}), is not the most general form of
the linear momentum in $j$ direction which satisfies the corresponding
relation as of (\ref{14}). By adding a general gauge function such
as $\psi\left(x,y,z\right)$ to (\ref{15}), (\ref{14}) still holds.
Hence, we propose 
\begin{equation}
P_{j}=-i\hbar\,g_{jj}^{^{-\frac{1}{2}}}\,\partial_{j}+\psi_{j}\,\label{16}
\end{equation}
in which $\psi_{j}=\psi_{j}\left(x,y,z\right)$ is an arbitrary gauge
function, corresponding to the $j$ direction. On the other hand,
$\hat{P}_{j}$ is supposed to represent the momentum operator in $j$
direction. This, however, suggests that $\hat{P}_{j}$ should be locally
Hermitian. To apply the local-Hermiticity condition, we assume that
the space is flat in the local coordinate system, spanned by the local
Cartesian coordinates $\left\{ x,y,z\right\} $, such that $\left(\partial_{j}\right)^{\dagger}=-\partial_{j}.$
Considering the latter, one finds $\psi_{j}=-\frac{1}{2}i\hbar\left(g_{jj}^{^{-\frac{1}{2}}}\right)_{,j}$and
consequently the locally Hermitian momentum operator becomes 
\begin{equation}
P_{j}=-\frac{1}{2}i\hbar\left(g_{jj}^{^{-\frac{1}{2}}}\,\partial_{j}+\partial_{j}g_{jj}^{^{-\frac{1}{2}}}\right)\,.\label{17}
\end{equation}
Staying in the local $\left\{ x,y,z\right\} $ coordinate system and
employing the corresponding local momentum operators (\ref{17}),
one, in principle, may write the time-independent Schrodinger equation
of a quantum particle, undergoing a general time-independent potential
$V\left(\overrightarrow{r}\right)$, as 
\begin{equation}
\frac{1}{2m}\left(P_{j}\right)^{2}\,\Psi\left(\overrightarrow{r}\right)+V\left(\overrightarrow{r}\right)\,\Psi\left(\overrightarrow{r}\right)=E\,\Psi\left(\overrightarrow{r}\right)\,,\label{18}
\end{equation}
in which $E$ and $m$ stand for the energy and the mass of the particle,
respectively, while $V\left(\overrightarrow{r}\right)$ is the time-independent
potential. Inserting (\ref{17}) in (\ref{18}), one explicitly finds{\small{}{}{}{}{}{}
\begin{multline}
-\frac{\hbar^{2}}{2m}\sum_{j=1}^{3}\Bigg(g_{_{,j}}^{jj}\,\partial_{j}+g^{jj}\,\partial_{j}^{2}+\frac{1}{4}\,g_{_{,j,j}}^{jj}-\frac{1}{16}\,g_{jj}\left(g_{_{,j}}^{jj}\right)^{2}\Bigg)\,\Psi\\
=(E-V)\,\Psi\,,\label{19}
\end{multline}
}in which $g_{_{,j}}^{jj}=\partial_{j}g^{jj}$ and $g_{_{,j,j}}^{jj}=\partial_{j}\partial_{j}g^{jj}$.
Our formalism also addresses the very recent work of Chung and Hassanabadi
\citep{Hassanabadi}, where the three-dimensional quantum mechanics
in a curved space, based on the $q$-addition, is studied. By looking
at the Eqs. (16)-(23) in Ref. \citep{Hassanabadi}, one observes that
their momentum operator suffers from the lack of Hermiticity \textcolor{black}{(see
the disscussion after Eq. (78) of Ref. \cite{LAVAGNO} ). This can
be seen from the definition of the $q$-momentum operator in Eq. (10)
of the Ref. \cite{Hassanabadi} where $\hat{p}=-i\hbar D_{x}$ or
equivalently 
\begin{equation}
\hat{p}=-i\hbar\frac{1}{1+qx}\partial_{x}.
\end{equation}
In terms of the canonical momentum, i.e., $\hat{p}_{0}=-i\hbar\partial_{x},$
it may be written as 
\begin{equation}
\hat{p}=\frac{1}{1+qx}\hat{p}_{0}.
\end{equation}
Knowing that the canonical momentum is Hermitian, with respect to
the standard definition, one finds 
\begin{equation}
\hat{p^{\dagger}}=\hat{p}_{0}\frac{1}{1+qx}=i\hbar\frac{q}{\left(1+qx\right)^{2}}+\hat{p}
\end{equation}
which is not clearly Hermitian i.e., $\hat{p^{\dagger}}\neq\hat{p}$
unless $q=0.$} In terms of the line element given in Ref. \citep{Hassanabadi}
(in Eqs. (20) and (21)), the corresponding Schrodinger equation (\ref{19})
reads 
\begin{multline}
-\frac{\hbar^{2}}{2m}\sum_{j=1}^{3}\Bigg((1+qx_{j})^{2}\,\partial_{j}^{2}+2q\left(1+qx_{j}\right)\,\partial_{j}+\frac{1}{4}q^{2}\Bigg)\,\Psi\\
=\left(E-V\right)\,\Psi\,.\label{20}
\end{multline}
(Note that here we use the convention $x_{j}=x,y,z$ for $j=1,2,3$,
respectively, while $\partial_{j}$ stands for $\dfrac{\partial}{\partial x_{j}}.$)
Continuing with the particle in a three-dimensional infinite box where
\begin{equation}
V\left(x,y,z\right)=\begin{cases}
0 & 0<x,y,z<L\\
\infty & \text{elsewhere}
\end{cases}\,,\label{21}
\end{equation}
the wave function is obtained to be

\begin{multline}
\Psi\left(x,y,z\right)=\frac{B}{\sqrt{\left(1+qx\right)\left(1+qy\right)\left(1+qz\right)}}\\
\times\sin\left[\frac{\ln\left(1+qx\right)}{\ln\left(1+qL\right)}n_{x}\pi\right]\,\sin\left[\frac{\ln\left(1+qy\right)}{\ln\left(1+qL\right)}n_{y}\pi\right]\\
\times\sin\left[\frac{\ln\left(1+qz\right)}{\ln\left(1+qL\right)}n_{z}\pi\right]\,,\label{22}
\end{multline}
with the same energy spectrum given in Eq. (29) of Ref. \citep{Hassanabadi}
and $B$ as a normalization constant. The nontrivial consequence of
the choice of $\hat{P}_{j}$ can be seen by comparing the eigenfunction
(\ref{22}) with (26) of the Ref. \citep{Hassanabadi}. Without giving
the details, one predicts the occurrence of similar differences for
the other examples in Ref. \citep{Hassanabadi}.

Here in this paper, parallel to the seminal work of Filho \textit{et
al.} \cite{Filho}, we first commence our work with the modification
of the quantum operators in terms of a background metric. When such
credible modification is found, we extend the metric-dependent Schrodinger
equation and Ehrenfest theorem. Going further, we will interpret our
results and then try to expand our findings in terms of the Poincare\ half-plane
metric for a free particle.

\section{Modification of quantum operators in a curved space}

To start our pursuit, we note that the internal product of two quantum
states on a Riemannian manifold can be written in the same way as
on the Euclidean geometry with a modified volume element that includes
the imprint of the background metric. We then write the internal product
of two quantum states, $\left|\chi\,\left(x_{1},x_{2},x_{3}\right)\right\rangle $
and $\left|\theta\,\left(x_{1},x_{2},x_{3}\right)\right\rangle $,
in a non-Euclidean geometry as {\small{}{}{}{}{}{} 
\begin{equation}
\left\langle \chi\mid\theta\right\rangle =\int\chi^{*}\,\left(x_{1},x_{2},x_{3}\right)\theta\,\left(x_{1},x_{2},x_{3}\right)\,\sqrt{\left|g\right|}\,dx_{1}\,dx_{2}\,dx_{3}\,,\label{23}
\end{equation}
}in which $x_{1}$, $x_{2}$, and $x_{3}$ are the local coordinates,
and the line element of the background space is introduced by $ds^{2}=g_{jk}dx^{j}dx^{k}$,
where $j,k=1,2,3$. Furthermore, $\left|g\right|$ is the absolute
value of the determinant of the metric tensor $g_{jk}$. For a diagonal
background metric, the modified differential element $dx_{j}$ can
be written as $g_{jj}^{-\frac{1}{2}}\,dx_{j}$ (no Einstein summation
rule) with $g_{jj}$ being the diagonal elements of the background
metric. Accordingly, the metric-dependent translation operator can
be written as \citep{Hassanabadi} 
\begin{equation}
T\,\left(g_{jj}^{-\frac{1}{2}}\left(x_{1},x_{2},x_{3}\right)\,dx^{j}\right)=1-\frac{i\,g_{jj}^{-\frac{1}{2}}\,p_{j}\,dx^{j}}{\hbar}\,,\label{24}
\end{equation}
and then 
\begin{equation}
T\left(\mathbf{\overrightarrow{x}}\right)=\exp\left(-\frac{i\,\overrightarrow{P}.\overrightarrow{x}}{\hbar}\right)\,.\label{25}
\end{equation}
In the above relation, \textbf{{}$\overrightarrow{P}$}{} is the
modified momentum operator in the presence of a non-flat diagonal
background metric. This new operator leads to the modified commutation
relation 
\begin{equation}
\left[x_{j},P_{k}\right]=i\hbar\,g_{kk}^{-\frac{1}{2}}\left(x_{1},x_{2},x_{3}\right)\,\delta_{jk}\,,\label{26}
\end{equation}
which clearly depends on the geometry. In other words, this commutation
is position-dependent and requires the modification of momentum operator
in the Euclidean space. The form of this altered momentum operator
could be naively considered as $P_{j}=-i\hbar\,g_{jj}^{-\frac{1}{2}}\partial_{j}$.
However, this form does not satisfy the Hermiticity of the new non-flat
momentum operator and thus needs to be modified as in (\ref{17}).
Note that, this Hermitian modified momentum operator obeys the same
commutation relation in (\ref{26}) since the augmented term $\partial_{j}g_{jj}^{-\frac{1}{2}}$
is only position-dependent and thus does not change the aforementioned
commutation relation. Now, we proceed with the injection of this new
Hermitian geometric momentum operator into the fundamental form-invariant
quantum equations. In this regard, the fact that the fundamental equations
of physics must be form-invariant under the change of the canonical
coordinates (due to the principle of Covariance \citep{kenneth}),
must be respected. Hence, the Hamiltonian function for a particle
with mass $m$ that moves in a non-Euclidean space can be expanded
with the new canonical variables $x_{j}$ and $P_{j}$ as 
\begin{equation}
H=\frac{1}{2m}\sum_{j}P_{j}^{2}+V\left(\mathbf{\overrightarrow{x}}\right)\,,\label{27}
\end{equation}
where one can expand $P_{j}^{2}$ operator with (\ref{17}) to obtain
\begin{multline}
P_{j}^{2}=-\frac{\hbar^{2}}{4}\Bigg[4\,g_{_{,j}}^{jj}\,\partial_{j}+g_{_{,j,j}}^{jj}+4\,g^{jj}\,\partial_{j}^{2}-\frac{1}{4}g_{jj}\big(g_{_{,j}}^{jj}\big)^{2}\Bigg]\,.\label{28}
\end{multline}
\textcolor{black}{By inserting (\ref{28}) into (\ref{27}), the following
relation for the three-dimensional modified Hamiltonian is obtained
which can be cast into the form (\ref{19})}

\begin{multline}
H=-\frac{\hbar^{2}}{2m}\sum_{j=1}^{3}\bigg[g^{jj}\partial_{j}^{2}+g_{_{,j}}^{jj}\partial_{j}+\frac{1}{4}g_{_{,j,j}}^{jj}\\
-\frac{1}{16}g_{jj}\left(g_{_{,j}}^{jj}\right)^{2}\bigg]+V(\mathbf{\overrightarrow{x}}).\label{29}
\end{multline}
Note that, all the terms of the Hamiltonian in a curved space are
dependent on the background metric in the absence of any imposed potential,
say for a free particle. Moreover, the Hermiticity of the modified
momentum operator, introduced by the last two terms in the brackets,
is position-dependent. In consequence, the commutation of the Hermitian
modified momentum operator with the Hamiltonian differs from the non-Hermitian
form of the modified momentum operator introduced before. This leads
to new terms in the Heisenberg equation for the new Hermitian canonical
coordinate $x_{j}$. In this regard, we explicitly see that the Hermiticity
condition of the modified momentum operator must be granted so one
can find a reliable behavior in a curved space.

\section{{\protect{\normalsize{}{}{}{}{}{}Ehrenfest theorem in a curved
space}}}

Imposing quantum mechanics on a curved space may add implications
that are worth probing. One of these implications is the additional
term in the Heisenberg uncertainty principle by which a non-zero minimal
momentum would arise \citep{mignemi,hamil,coelho}. In fact, if a
background metric is a smooth function of the class $C^{\infty}$,
derivatives of the metric from any order exist and are continuous.
Thus, Taylor-series expansion of this metric leads to the emergence
of a quadratic term $x_{j}^{2}$ in the Heisenberg uncertainty principle,
if the expansion is truncated at the second order. Consequently, this
modification will cause a non-zero minimal energy in any quantum system
that is being studied in a curved space. Looking more carefully, the
emergence of a non-zero minimal energy for a quantum system in a curved
space is consistent with the notion of general relativity that implies
the connection between energy-momentum tensor and metric of the space.
Furthermore, this modification underlines the non-commutativity and
discretization of the momentum space \citep{mignemi}. This can be
expressed as 
\begin{equation}
\Delta x_{j}\,\Delta P_{k}\geq\frac{\hbar}{2}\left[1+\alpha\,\left(\Delta x_{k}^{2}+\left\langle x_{k}\right\rangle ^{2}\right)\right]\delta_{jk}\,,\label{30}
\end{equation}
in which $\alpha$ is a positive arbitrary constant that causes significant
modifications in the very large scales. \textcolor{black}{In the line
of the descriptions above, $\alpha$ would be relevant to the emergent
minimal energy of a quantum system. This can be easily interpreted
for the case $\left\langle x_{k}\right\rangle \,=\,0$, by choosing
the appropriate coordinate system, as writing Eq. (\ref{30}) in the
equality limit. In this way, the minimal measurable momentum is calculated
as $\hbar\sqrt{\alpha}$, implying the physical meaning of $\alpha$
in the Eq. (\ref{30}). There have been so far some theoretical efforts
to constrain the value of $\alpha$ in the literature \citep{Mirza},
but none of them could be well linked to the phenomenology in an experimental
manner.}

The modified momentum operator in a curved space can be further investigated
in the Heisenberg picture. One may try to examine any likely revamp
of the Ehrenfest theorem in a curved space assuming that the Heisenberg
equation is still valid in a non-flat space. For a particle with mass
$m$, it is found that

\begin{equation}
\frac{d\left\langle x_{j}\right\rangle }{dt}=\frac{1}{i\hbar}\left\langle \left[x_{j},H\right]\right\rangle =\frac{1}{2m}\left\langle \left\{ g_{jj}^{-\frac{1}{2}},P_{j}\right\} \right\rangle ,\label{31}
\end{equation}
and 
\begin{equation}
\frac{d\left\langle P_{k}\right\rangle }{dt}=\frac{1}{i\hbar}\left\langle \left[P_{k},H\right]\right\rangle =-\left\langle g_{kk}^{-\frac{1}{2}}\,\partial_{k}V\left(\mathbf{\overrightarrow{x}}\right)\right\rangle \,,\label{32}
\end{equation}
where $\left\{ \right\} $ denotes anti-commutation. Mathematical
manipulation of (\ref{32}) leads to another useful relation given
by 
\begin{equation}
\frac{d\left\langle P_{k}\right\rangle }{dt}=\frac{1}{2}\left\langle g_{kk}^{-\frac{1}{2}}\,\frac{d}{dt}\left\{ g_{kk}^{-\frac{1}{2}},P_{k}\right\} \right\rangle \,.\label{33}
\end{equation}
Using Eqs. (\ref{32}) and (\ref{33}), one can write the relation
\begin{multline}
\frac{d\left\langle P_{k}\right\rangle }{dt}=\\
-\frac{1}{2}\left\langle g_{kk}^{-\frac{1}{2}}\,\partial_{k}V\left(\mathbf{\overrightarrow{x}}\right)\right\rangle +\frac{1}{4}\frac{d}{dt}\left\langle g_{kk}^{-\frac{1}{2}}\left\{ g_{kk}^{-\frac{1}{2}},P_{k}\right\} \right\rangle \label{34}
\end{multline}
in the Heisenberg picture. We would like to add that a specific form
of the latter relation has been found in \cite{Filho} (Eq. (11))
where $\left\langle p_{x}\right\rangle ,$ in the lhs, stands for
$m\left\langle \dot{x}\right\rangle $. The term $g_{kk}^{-\frac{1}{2}}\left\{ g_{kk}^{-\frac{1}{2}},P_{k}\right\} $
can be further simplified using 
\begin{equation}
\left[P_{k},g_{kk}^{-\frac{1}{2}}\right]=-i\hbar\,g_{kk}^{-\frac{1}{2}}\,g_{kk\,,k}^{-\frac{1}{2}}\,,\label{35}
\end{equation}
to obtain 
\begin{equation}
g_{kk}^{-\frac{1}{2}}\left\{ g_{kk}^{-\frac{1}{2}},P_{k}\right\} =2g^{kk}\,P_{k}-\frac{i\hbar}{2}g_{kk}^{-\frac{1}{2}}\,g_{,k}^{kk}\,,\label{36}
\end{equation}
where the second term on the right-hand side depends only on the position.
Finally, Eq. (\ref{34}) can be expressed as 
\begin{equation}
\frac{d\left\langle P_{k}\right\rangle }{dt}=-\frac{1}{2}\left\langle g_{kk}^{-\frac{1}{2}}\,\partial_{k}V\left(\mathbf{\overrightarrow{x}}\right)\right\rangle +\frac{1}{2}\frac{d}{dt}\left\langle g^{kk}\,P_{k}\right\rangle \,,\label{37}
\end{equation}
assuming that the background metric is \textit{static}. The latter
equation can be considered as the modified Ehrenfest theorem in a
curved space. Evidently, the choice $g_{kk}=1$ reduces this equation
to its original form in Euclidean space, given by 
\begin{equation}
\frac{d\left\langle P_{k}\right\rangle }{dt}=-\left\langle \partial_{k}V(\mathbf{\overrightarrow{x}})\right\rangle \,.\label{38}
\end{equation}
It must be noted that the Ehrenfest theorem was acquired from the
Heisenberg equation with the assumption that the Heisenberg equation
is form-invariant in a curved space, comparing with a flat geometry.
In consequence, it is reasonable to expect that the Ehrenfest theorem
is form-invariant in a curved space, as well. As it was mentioned,
Eq. (\ref{37}) can be looked at as a modification of the Ehrenfest
theorem in a curved space. However, the same flat form for the Ehrenfest
theorem can be deduced considering a new geometric position-dependent
differential operator $D_{k}\equiv\frac{1}{2}g_{kk}^{-\frac{1}{2}}\,\partial_{k}$,
along with an effective potential that includes the imprint of the
curved space \citep{kanat}. This effective potential comprises an
external imposed potential as well as tidal forces due to the background
geometry. Thus, for a static background metric in the Heisenberg picture,
Eq. (\ref{37}) can be written in the form 
\begin{equation}
\frac{d\left\langle P_{k}\right\rangle }{dt}=-\bigg\langle\text{\ensuremath{D_{k}\big(V_{\text{ext}}\left(\mathbf{\overrightarrow{x}}\right)+V_{\text{curv}}\left(\overrightarrow{x}\right)\big)}}\bigg\rangle\,,\label{39}
\end{equation}
in which $D_{k}V_{\text{curv}}\left(\overrightarrow{x}\right)\equiv-\frac{1}{2}\frac{d}{dt}\left(g^{kk}P_{k}\right)$.
In this fashion, the invariant form of the Ehrenfest theorem in a
curved space with the modified differential operator $D_{k}$ and
the mapped effective potential $V_{\text{eff}}\left(\mathbf{\overrightarrow{x}}\right)\equiv V_{\text{ext}}\left(\overrightarrow{x}\right)+V_{\text{curv}}\left(\overrightarrow{x}\right)$
can be written as 
\begin{equation}
\frac{d\left\langle P_{k}\right\rangle }{dt}=-\left\langle D_{k}V_{\text{eff}}\left(\overrightarrow{x}\right)\right\rangle \,.\label{40}
\end{equation}
For instance, we may consider a free particle on which no external
potential is imposed. Thus, Eq. (\ref{37}) reads 
\begin{equation}
\frac{d\left\langle P_{k}\right\rangle }{dt}=\frac{1}{2}\frac{d}{dt}\left\langle g^{kk}\,P_{k}\right\rangle \,,\label{41}
\end{equation}
that non-trivially indicates a non-zero average force acting on a
free particle due to the curved space. From the above discussion,
it is obvious that the average force imposed on the free particle
vanishes when the background geometry is flat. Hence, no quantum particle
can be considered free from exterior potentials in a curved space,
contrary to the Euclidean case. It is interesting to note that the
geometry of the space does not impose any average force on a quantum
system in case of a flat geometry. This implication can be concluded
from Eq. (\ref{41}), where a flat metric results in

\begin{equation}
\frac{d\big<P_{k}\big>}{dt}=0\,.\label{42}
\end{equation}
This outcome explicitly shows that only a spatially changing geometry
of the space is able to impose an effective average force on a quantum
system in view of the modification of quantum mechanics in a curved
space.

\section{{\protect{\normalsize{}{}{}{}{}{}A free particle on the Poincare\ half-plane
geometry}}}

We proceed with investigating the behavior of a quantum particle free
from any external potentials in the Poincare upper half-plane geometry.
This represents a two-dimensional hyperbolic upper half-plane geometry
\citep{terras} that imposes an effective constraining force on the
particle, based on our above-mentioned results. The metric for the
upper half-plane $x$ is defined as

\begin{equation}
ds^{2}=\frac{dx^{2}+dy^{2}}{x^{2}}\,,\quad x>0\,,\label{43}
\end{equation}
which has no dependence on dimension $z$. Hence, the wave function
of the particle will be independent of $z$ according to Eq. (\ref{8}),
considering $V\left(x,y,z\right)=0$.

The modified time-independent Schrodinger equation $H\,\Psi\left(x,y\right)=E\,\Psi\left(x,y\right)$
in the new geometry is explicitly given by 
\begin{multline}
-\frac{\hbar^{2}}{2m}\Bigg[x^{2}\Big(\partial_{x}^{2}\Psi\left(x,y\right)+\partial_{y}^{2}\Psi\left(x,y\right)\Big)\\
+2x\,\partial_{x}\Psi\left(x,y\right)+\frac{1}{4}\Psi\left(x,y\right)\Bigg]=E\,\Psi\left(x,y\right)\,,\quad x>0\,,\label{44}
\end{multline}
in accordance with the Hamiltonian (\ref{8}) and the line element
(\ref{43}). To solve Eq. (\ref{44}), we assume that $\Psi(x,y)$
is a separable function with respect to its arguments, i.e. $\Psi\left(x,y\right)=\phi\left(x\right)\,\theta\left(y\right)$,
noting the commutativity of the Poincare\ geometry. Thus Eq. (\ref{44})
reads as 
\begin{multline}
-\frac{\hbar^{2}}{2m}\Bigg[\frac{1}{\phi\left(x\right)}\,\frac{d^{2}\phi\left(x\right)}{dx^{2}}+\frac{2}{x\,\phi\left(x\right)}\,\frac{d\phi\left(x\right)}{dx}\\
+\frac{1}{4\,x^{2}}+\frac{1}{\theta\left(y\right)}\,\frac{d^{2}\theta\left(y\right)}{dy^{2}}\Bigg]=\frac{E}{x^{2}}\quad,\,x>0\,,\label{45}
\end{multline}
which leads to separate equations in the set 
\begin{equation}
\left\{ \begin{array}{l}
\frac{1}{\phi\left(x\right)}\,\frac{d^{2}\phi\left(x\right)}{dx^{2}}+\frac{2}{x\,\phi\left(x\right)}\,\frac{d\phi\left(x\right)}{dx}+\frac{\frac{1}{4}+\frac{2mE}{\hbar^{2}}}{x^{2}}=A\\
\\
\frac{1}{\theta\left(y\right)}\frac{d^{2}\theta\left(y\right)}{dy^{2}}=-A
\end{array}\right.\,,\label{46}
\end{equation}
for $x>0$ with $A$ being an arbitrary constant. The latter equation
for $\theta\left(y\right)$ is similar to its counterpart for a free
particle in the Euclidean space and has the solution 
\begin{equation}
\theta\left(y\right)=C_{1}\,e^{ik\,y}+C_{2}\,e^{-ik\,y}\,,\label{47}
\end{equation}
where $k\equiv\sqrt{A}$ is the spatial frequency of the particle,
while $C_{1}$ and $C_{2}$ are integration constants. The solution
to the first equation in (\ref{46}) is of type Bessel function. To
find the exact solution, one can transform it into a standard Schrodinger
equation by a point canonical transformation \cite{pct} (and references
therein). To this end, we apply the transformation 
\begin{equation}
\phi\left(x\right)=\frac{1}{\sqrt{x}}\,\zeta\left(z\left(x\right)\right),\label{54}
\end{equation}
where

\begin{equation}
z\left(x\right)=\ln\left(x\right)\,,\label{55}
\end{equation}
to obtain 
\begin{equation}
-\frac{\hbar^{2}}{2m}\!\left(\frac{d^{2}\zeta\left(z\right)}{dz^{2}}\right)+V_{\mathrm{eff}}\left(z\right)\,\zeta\left(z\right)=E\,\zeta\left(z\right)\,,\label{56}
\end{equation}
with 
\begin{equation}
V_{\mathrm{eff}}\left(z\right)=\frac{k^{2}\hbar^{2}}{2m}\,e^{2z}\,.\label{57}
\end{equation}

\begin{figure}[h]
\includegraphics[width=6cm]{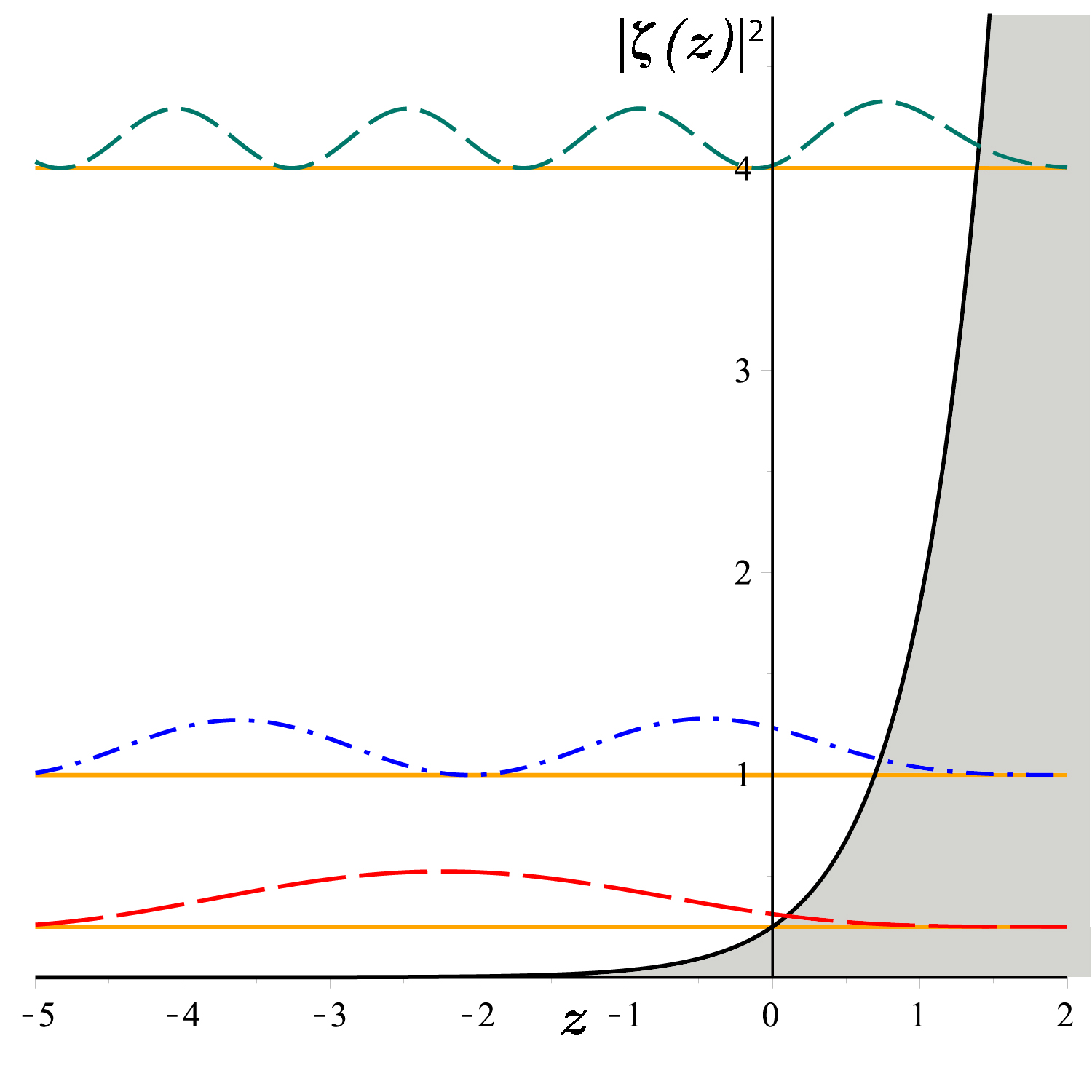} \caption{Non-normalized $\left|\zeta\left(z\right)\right|^{2}$ in terms of
$z$ for $\omega=\frac{1}{2}$ (\textcolor{red}{Red}-Long Dash), $1$
(\textcolor{blue}{Blue}-Dash Dot) and $2$ ( \textcolor{green}{Green}-Dash)
with $k=\frac{1}{2}.$ The scaled effective potential i.e., $V_{\mathrm{eff}}\left(z\right)/\frac{k^{2}\hbar^{2}}{2m}$
is also plotted in the solid (black) curve. The plots clearly depict
the expected asymptotic behaviors when $z\rightarrow\pm\infty.$ The
horizontal lines stand for the scaled energy i.e. $\omega^{2}\equiv\frac{2mE}{\hbar^{2}}$.}
\label{Fig. 1} 
\end{figure}

Introducing $\omega^{2}\equiv\frac{2mE}{\hbar^{2}}$, one finds a
solution for the Schrodinger equation (\ref{56}), in terms of the
modified Bessel functions $I_{\nu}$ and $K_{\nu}$ of the first and
the second kind, respectively. This solution is given by 
\begin{equation}
\zeta\left(z\right)=D_{1}\,I_{i\omega}\left(ke^{z}\right)+D_{2}\,K_{i\omega}\left(ke^{z}\right)\label{58}
\end{equation}
in which $D_{1}$ and $D_{2}$ are integration constants. Due to the
asymptotic behavior of the effective potential, $V_{\mathrm{eff}}\left(z\right)\sim\,e^{2z}$,
one expects the solution to be of the form of a plane wave for $z\rightarrow-\infty,$
and zero when $z\rightarrow\infty.$ Hence, one must set $D_{1}=0$
which implies 
\begin{equation}
\zeta\left(z\right)\sim K_{i\omega}\left(ke^{z}\right)\label{58}
\end{equation}
with a continuous positive energy. In FIG. \ref{Fig. 1} we plot the
non-normalized $\left|\zeta\left(z\right)\right|^{2}$ in terms of
$z$ for $\omega=\frac{1}{2},1$ and $2$ and $k=\frac{1}{2}.$ The
scaled effective potential and energy are also displayed.

\begin{figure}[h]
\centering\includegraphics[width=6cm]{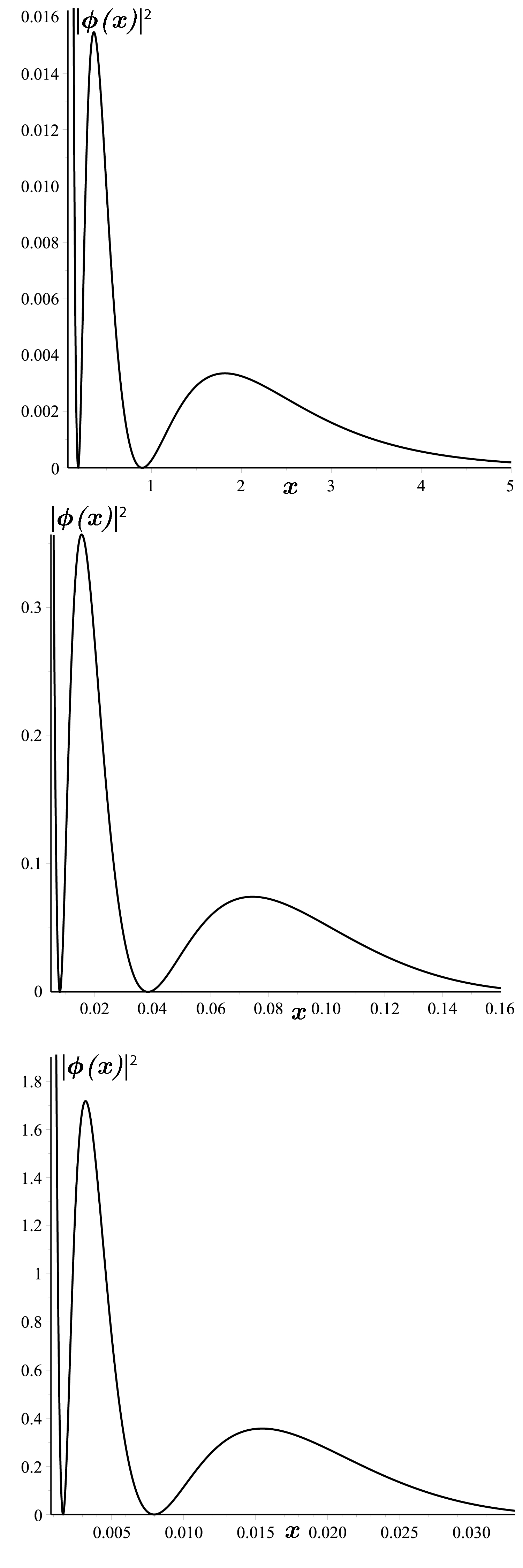} \textcolor{black}{\caption{Plots of $\left|\phi\left(x\right)\right|^{2}$ in terms of $x$ for
$\omega=2$ and $k=\frac{1}{2}$ for three different intervals. \textcolor{black}{The
self-similarity of the probability densities is in agreement with
the classical geodesics of the Poincare\ upper half-plane geometry
(see the Appendix A). }}
\label{Fig. 2} }
\end{figure}

The wave function in $x$ direction i.e., $\phi\left(x\right)$ is
obtained through an inverse transformation and expressed as 
\begin{equation}
\phi\left(x\right)\sim\frac{1}{\sqrt{x}}K_{i\omega}\left(kx\right)\label{58}
\end{equation}
up to a normalization constant. In Fig. \ref{Fig. 2} we plot $\left|\phi\left(x\right)\right|^{2}$
in terms of $x$ for $\omega=2$ and $k=\frac{1}{2}$ for three different
intervals of $x$. \textcolor{black}{This figure implies the self-similarity
symmetry of the probability density for $x>0$. This is in agreement
with the classical geodesics of the Poincare\ upper half-plane geometry
which is scale-invariant / self-similar (see the Appendix A) .}

\textcolor{black}{We are also interested to inspect the energy spectrum
in the case of very low effective potential, where $x$ tends to zero.
This is being done for the sake of finding any clue for the emergence
of non-zero minimal energy in the spectrum of a free particle, as
placed on the upper half-plane geometry. In this way, we note that
the second derivative of $\zeta\left(z\right)$ becomes zero in the
mentioned limit, while the $\zeta\left(z\right)$ itself becomes indefinite.
We then write Eq. (\ref{56}) in the following form 
\begin{equation}
-\frac{\hbar^{2}}{2m}\!\left(\frac{d^{2}\zeta\left(z\right)}{dz^{2}}\right)=\left(E-V_{\mathrm{eff}}\right)\,\zeta\left(z\right)\,,\label{59}
\end{equation}
and interpret that the term $E-V_{\mathrm{eff}}$ must be equal to
zero in the limit $x\rightarrow0$. So, energy would be equal to the
effective potential in the limit that the particle is asymptotically
behaving as a plane wave. In advance, a minimal effective potential
due to the curvature of the background, look at Eq. (\ref{41}), and
also due to accepting the minimal length cutoff (so that $|z|$ marks
a maximum value in our case), can be regarded. In this manner we have
shown that the claim of a minimal non-zero energy for a free quantum
particle which is placed on the upper half-plane background can be
confirmed.}

\section{{\protect{\normalsize{}{}{}{}{}{}Conclusion}}}

To migrate from the quantum mechanics on the Euclidean space to its
extension on a curved space, one should include the metric-dependent
terms in the relevant generalized equations. In this regard, our geometrical
approach in this study was initiated by designating a modified momentum
operator such that it conveys the geometry of a curved space while
engrossing Hermiticity. By this definition, the extraction of the
corresponding Hamiltonian and the interpretation of the generalized
Ehrenfest theorem followed. It was observed that Hamiltonian depends
on the components of the background curved space, as well as the position
of the particle. This, in turn, leads to emergence of new terms in
the Heisenberg equation and modifies the Ehrenfest theorem (Eq. \ref{37}).
However, it was shown that the Ehrenfest theorem could be put in an
invariant form using a new geometric position-dependent differential
operator $D_{k}\equiv\frac{1}{2}g_{kk}^{-\frac{1}{2}}\,\partial_{k}$
and an effective potential that includes the tidal forces due to the
curved background geometry. More importantly, it can be understood
from (\ref{41}) that flat space imposes no constraint on a free particle,
although it is not true for a general geometry.

In continuation, we derived the wave function of a ``free particle''
moving in the Poincare\ upper half-plane geometry. It was observed
that the ``free particle'' that is indeed free along the $y$-axis,
actually behaves as if it is bounded by the curved space along the
$x$-axis (FIG. \ref{Fig. 1}), due to the term $\frac{1}{\sqrt{x}}$
in $\phi(x)$. In fact, there exists an effective potential barrier
that surges exponentially towards infinity as $x\rightarrow+\infty$
. Thus, the probability density $\left|\phi\left(x\right)\right|^{2}$
approaches zero while $x$ increases towards infinity. This, of course,
is explicable due to the Poincare\ half-plane geometry, for which
the geodesics of a classical free particle are also bounded. In addition,
we should notice the behavior of the particle while $x\rightarrow0$.
In FIG. \ref{Fig. 2}, the probability density shows a fractal character
in the vicinity of $x=0$, added by its increase while moving towards
$x=0$. In fact, effective potential becomes zero while $x\rightarrow0$.
In consequence, we expect the oscillatory behavior of a Euclidean
free particle as it is but with the increasing probability density
as $x\rightarrow0$. To recapitulate, a particle free from any exterior
potential that is put in the Poincare\ upper half-plane geometry
experiences a bounded behavior in accord with its classical fractal
geodesics.

\appendix

\section{\textcolor{black}{Classical Geodesics in the Poincare Upper Half
Plane}}

\textcolor{black}{Considering the line-element (\ref{43}), the classical
geodesics equation in the Poincare upper half-plane is given by 
\begin{equation}
\frac{d^{2}x^{i}}{d\sigma^{2}}+\Gamma_{jk}^{i}\frac{dx^{j}}{d\sigma}\frac{dx^{k}}{d\sigma}=0\tag{A1}\label{A1}
\end{equation}
in which $\sigma$ is the geodesic parameter and $\Gamma_{jk}^{i}$
are the Christoffel symbols. Knowing that the only nonzero Christoffel
symbols are $\Gamma_{xx}^{x}=-\Gamma_{yy}^{x}=-\Gamma_{xx}^{y}=\frac{1}{x}$,
one finds the following equations 
\begin{equation}
\ddot{x}-\frac{1}{x}\dot{x}^{2}+\frac{1}{x}\dot{y}^{2}=0\tag{A2}\label{A2}
\end{equation}
and 
\begin{equation}
\ddot{y}-\frac{2}{x}\dot{x}\dot{y}=0\tag{A3}\label{A3}
\end{equation}
in which a dot stands for the derivative with respect to $\sigma$.
Eq. (\ref{A3}) yields 
\begin{equation}
\dot{y}=\frac{\dot{y}_{0}}{x_{0}^{2}}x^{2}\tag{A4}\label{A4}
\end{equation}
and consequently Eq. (\ref{A2}) admits 
\begin{equation}
\dot{x}=x\left(\frac{\dot{x_{0}}+\dot{y}_{0}}{x_{0}}-\frac{\dot{y}_{0}}{\dot{x}_{0}^{2}}y\right).\tag{A5}\label{A5}
\end{equation}
For, $\dot{y}_{0}=0,$ the geodesic is a straight line parallel to
the $x$-axis. With $\dot{y}_{0}\neq0$, the combination of (\ref{A4})
and (\ref{A5}) gives 
\begin{equation}
\frac{dy}{dx}=\frac{\frac{\dot{y}_{0}}{x_{0}^{2}}x}{\left(\frac{\dot{x_{0}}+\dot{y}_{0}}{x_{0}}-\frac{\dot{y}_{0}}{\dot{x}_{0}^{2}}y\right)}\tag{A6}.\label{A6}
\end{equation}
The latter first order ODE admits exact solution given by 
\begin{equation}
x^{2}+\left(y-y_{c}\right)^{2}=R_{c}^{2}\tag{A7}\label{A7}
\end{equation}
which is the geodesic of the classical particle. Herein, $y_{c}=x_{0}\left(1+\frac{\dot{x}_{0}}{\dot{y}_{0}}\right)$
and $R_{c}^{2}=x_{0}^{2}+\left(y_{0}-y_{c}\right)^{2}$ and the curve
is a semi-circle centered at $\left(0,y_{c}\right)$ with radius $R_{c}.$
It is worth mentioning that, the geodesic in Poincare upper half plane
is self-similar or scale-invariant. To see this let's introduce the
transformation $x=\lambda\widetilde{x}$ and $y=\lambda\widetilde{y}$
for some scale factor $\lambda.$ Upon this transformation, Eq. (\ref{A7})
becomes 
\begin{equation}
\widetilde{x}^{2}+\left(\widetilde{y}-\frac{y_{c}}{\lambda}\right)^{2}=\frac{R_{c}^{2}}{\lambda^{2}}\tag{A8}\label{A8}
\end{equation}
which is again a semi circle with scaled radius $\widetilde{R}_{c}$,
centered at $\left(0,\widetilde{y}_{c}\right)$ in which $y_{c}=\lambda\widetilde{y}_{c}$
and $R_{c}=\lambda\widetilde{R}_{c}$ with the same scale factor. }
\end{document}